\documentclass[fleqn,usenatbib]{mnras}

\usepackage{newtxtext,newtxmath}
\usepackage{float}
\usepackage[utf8]{inputenc}
\usepackage[T1]{fontenc}
\usepackage{ae,aecompl}
\usepackage{subcaption}
\usepackage[colorinlistoftodos,prependcaption,textsize=tiny]{todonotes}

\usepackage{graphicx}	
\usepackage{amsmath}	
\usepackage{amssymb}	
\usepackage{color}
\usepackage[]{units}
\usepackage{titlesec}

\titlespacing{\section}{0pt}{\baselineskip}{-\parskip}

\title[Primordial gas clouds]{The effects of metallicity and cooling physics on fragmentation: implications on direct-collapse black hole formation} 

\author[C. Corbett Moran et al.]{
C. Corbett Moran,$^{1}$\thanks{E-mail: corbett@caltech.edu}
M. Y. Grudi\'c,$^{1}$
P. F. Hopkins$^{1}$
\\
$^{1}$ TAPIR, Mailcode 350-17, California Institute of Technology, Pasadena, CA 91125, USA
}

\date{Accepted XXX. Received YYY; in original form ZZZ}

\pubyear{2017}

\begin{document}
\label{firstpage}
\pagerange{\pageref{firstpage}--\pageref{lastpage}}
\maketitle

\begin{abstract}
A promising supermassive black hole seed formation channel is that of direct collapse from primordial gas clouds. We  perform a suite of 3D hydrodynamics simulations of an isolated turbulent gas cloud to investigate conditions conducive to forming massive black hole seeds via direct collapse, probing the impact of cloud metallicity, gas temperature floor and cooling physics on cloud fragmentation.  We find there is no threshold in metallicity which produces a sharp drop in fragmentation. When molecular cooling is not present, metallicity has little effect on fragmentation. When molecular cooling is present, fragmentation is suppressed by at most $\sim 25\%$, with the greatest suppression seen at metallicities below $2\%$ solar. A gas temperature floor $\sim 10^{4}$K produces the largest drop in fragmentation of any parameter choice, reducing fragmentation by $\sim 60\%$.  At metallicities below $2\%$ solar or at temperatures $\sim 10^{3}$K we see a reduction in fragmentation $\sim 20-25 \%$. For a cloud of metallicity $2\%$ solar above and a temperature below $10^3$K, the detailed choices of temperature floor, metallicity, and cooling physics have little impact on fragmentation. 
\end{abstract}



\section{INTRODUCTION}
Supermassive black holes are in place by the first billion years of the universe's existence \citep{willott_four_2007,mortlock_luminous_2011,wu_ultraluminous_2015,venemans_identification_2015,jiang_discovery_2015,wang_first_2017}. Several promising channels have be proposed for their formation, but forming supermassive black holes in the requisite time-frame remains a theoretical puzzle.

In the Population III scenario, a Population III star collapses to a seed black hole $\sim 10-100 M_\odot$ \citep{clark_gravitational_2011}, which then grows to supermassive size via rapid accretion \citep{volonteri_assembly_2003,madau_massive_2001}. \mbox{\cite{alexander_rapid_2014}} find that a Pop III remnant of typical size $\sim 10-100 M_\odot$ \citep{clark_gravitational_2011,stacy_first_2012} can grow to a massive $10^4 M_\odot$ seed in $\sim 10^7$ years, assuming a model of supra-exponential accretion. The enormous accretion rate that must be maintained for this scenario to be plausible motivates considering alternatives. 

One promising such scenario is direct collapse, in which a cloud of gas directly collapses to a massive seed black hole  $M_{BH}\approx 10^{4-5} M_\odot$ \citep{eisenstein_origin_1995,oh_second-generation_2002,bromm_formation_2003,koushiappas_massive_2004,lodato_supermassive_2006}. The massive seed black hole then grows by accretion, but unlike in the Population III scenario, this accretion can comfortably proceed at sub-Eddington rates. The collapsed object is able to retain up to 90\% of the the initial clouds' Jeans mass \citep{begelman_formation_2006,regan_pathways_2009,latif_black_2013,choi_supermassive_2013} under the assumption that atomic hydrogen be the only coolant available in the cloud \citep{volonteri_formation_2010,natarajan_formation_2011,haiman_formation_2013} due to an external UV source in the region suppressing molecular hydrogen formation \citep{dijkstra_fluctuations_2008,agarwal_first_2014}.

The primary obstacle to the production of a massive direct-collapse object is fragmentation of the lower density progenitor cloud under the gravitational instability, which would cause the gas supply to be exhausted forming a cluster of less-massive clumps or stars rather than a single mass object. It is well known that fragmentation depends centrally on the ratio of cooling to collapse timescales. ISM cooling is generally faster at higher metallicity, as the presence of dust and metals enables a wide variety of cooling mechanisms that can ultimately cool the ISM to as low as $\unit[\sim 10]{K}$. Therefore, a more metal-rich ISM should fragment more readily, with all other things being equal. 

In this work we focus on simulations of conditions conducive to forming massive seed black holes via direct collapse. We perform a suite of 3D hydrodynamical simulations of primordial gas clouds to explore the relative contribution of each of these parameters to inhibiting fragmentation, as well as to comfortably exclude regions of parameter space which have no impact on fragmentation. In \textbf{Section ~\ref{sec:sims}}, we list the methods used for our numerical models. In \textbf{Section ~\ref{sec:results}} we present our results for fragmentation as a function of metallicity,  and the temperature of the ISM. In \textbf{Section ~\ref{sec:conclusions}} we present our conclusions.
\section{SIMULATIONS}\label{sec:sims} 
Our suite of 3D hydrodynamical simulations uses GIZMO \citep{hopkins_new_2015}, a Lagrangian mesh-free finite-volume Godunov code. GIZMO is built upon the gravity and domain algorithms within GADGET-3 \citep{springel_cosmological_2005} with the advantages of both smoothed-particle hydrodynamics and grid-based models. Simulations are run in GIZMO's Meshless-Finite Mass mode. 

Simulations labeled GIZMO use the star formation and feedback models of the FIRE project \citep{hopkins_galaxies_2014,hopkins_stellar_2012}. These included metal-line, Compton, photo-ionization, photo-electric, molecular and fine-structure processes. Optically thick cooling is self consistently accounted for including dust-gas cooling as in \cite{hopkins_galaxies_2014}. Simulations labeled molecular (H2) or molecular/deuterium (H2+D) additionally incorporate the \texttt{Grackle} library for non-equilibrium chemistry and cooling \citep{smith_grackle:_2017}. \texttt{Grackle} is used in both a 9 species (molecular) and 12 species network configuration (molecular/deuterium). The molecular (H2) configuration tracks abundances of H, H$^+$, He, He$^+$, He$^{++}$, e$^-$, H$_2$, H$^-$, and H$_2^+$, the molecular/deuterium (H2+D) configuration adds to this model D, D$^+$ and HD. While GIZMO cooling includes dust-gas coupling, \texttt{Grackle} does not; however, as this study is focused on metallicities below 10\% solar, dust cooling is not an appreciable cooling channel. In both models, we enforce a temperature floor (lower limit) equal to $\lfloor T \rfloor$.

Our prescription for star formation is a modified version of that used in \citet{hopkins_galaxies_2014}, and identical to \citet{grudic_when_2018}. A local virial parameter $\alpha$ \citep{hopkins_meaning_2013} is evaluated for each gas particle including thermal and turbulent contributions. If $\alpha < 1$ then the particle is considered self-gravitating and it is eligible to turn into a star particle, with a probability per unit time of $\epsilon_{ff} f_{mol} t_{ff}^{-1}$, where $\epsilon_{ff}$ is the local per-freefall SFE parameter (1 by default), $f_{mol}$ is the local molecular mass fraction of the gas, and $t_{ff}=\sqrt{\frac{3\pi}{32 G \rho}}$ is the local freefall time.

Star particles in the simulations inject mass, energy, and momentum into the ISM via radiation pressure, stellar winds, supernovae, photo-ionization and photo-electric heating. IMF-averaged properties are assigned to all star particles. Since each star is sampling a distribution, all star particles have the same mass. Thus the mass in stars is directly proportional to the number of stars formed in simulation. Initial conditions are set to be constant density gas sphere of radius 5 pc and mass $10^6 M_\odot$. The initial velocity field is set as in \cite{grudic_when_2018}, as a superposition of solid-body rotation about origin in conjunction with a component with random turbulence with a virial parameter of order unity (see David Guszejnov, in preparation for detailed study of initial turbulence field effects on fragmentation). The rotational frequency is $\Omega_K=\sqrt{\frac{R^3}{GM}}$. Simulations are run for 1 Myr.

\section{RESULTS}
\label{sec:results}
To measure fragmentation we define $M_{cluster}$ as the mass of the most massive cluster of star particles at 1 Myr  and normalize it by the initial mass of the gas, $M_{gas,init}$.  The higher the value, the more mass has accreted onto a single object, and thus the less fragmentation has occurred. $M_{cluster}$ is determined by running the cluster finding algorithm HOP2 \citep{grudic_top_2017}.  As a baseline for percentage comparison for all results, we consider metallicity of $0.1 Z/Z_\odot$, $\lfloor T \rfloor=10$K, $\epsilon_{ff}= 1.0$, and GIZMO cooling physics. 

\begin{figure}
\centering
\caption{Simulations of fragmentation of a dense, metal-poor gas cloud with fixed $\lfloor T \rfloor=10$K and $\epsilon_{ff}= 1.0$ and varied cooling physics, metallicity, and resolution. The mass of the most massive cluster of star particles at 1 Myr $M_{cluster}	$ is normalized by the initial mass of the gas, $M_{gas,init}$ and plotted as a function of metallicity. The effect of varying metallicity and cooling physics is largely independent of resolution.}
\begin{subfigure}[b]{0.4\textwidth}
	\includegraphics[width=\textwidth]{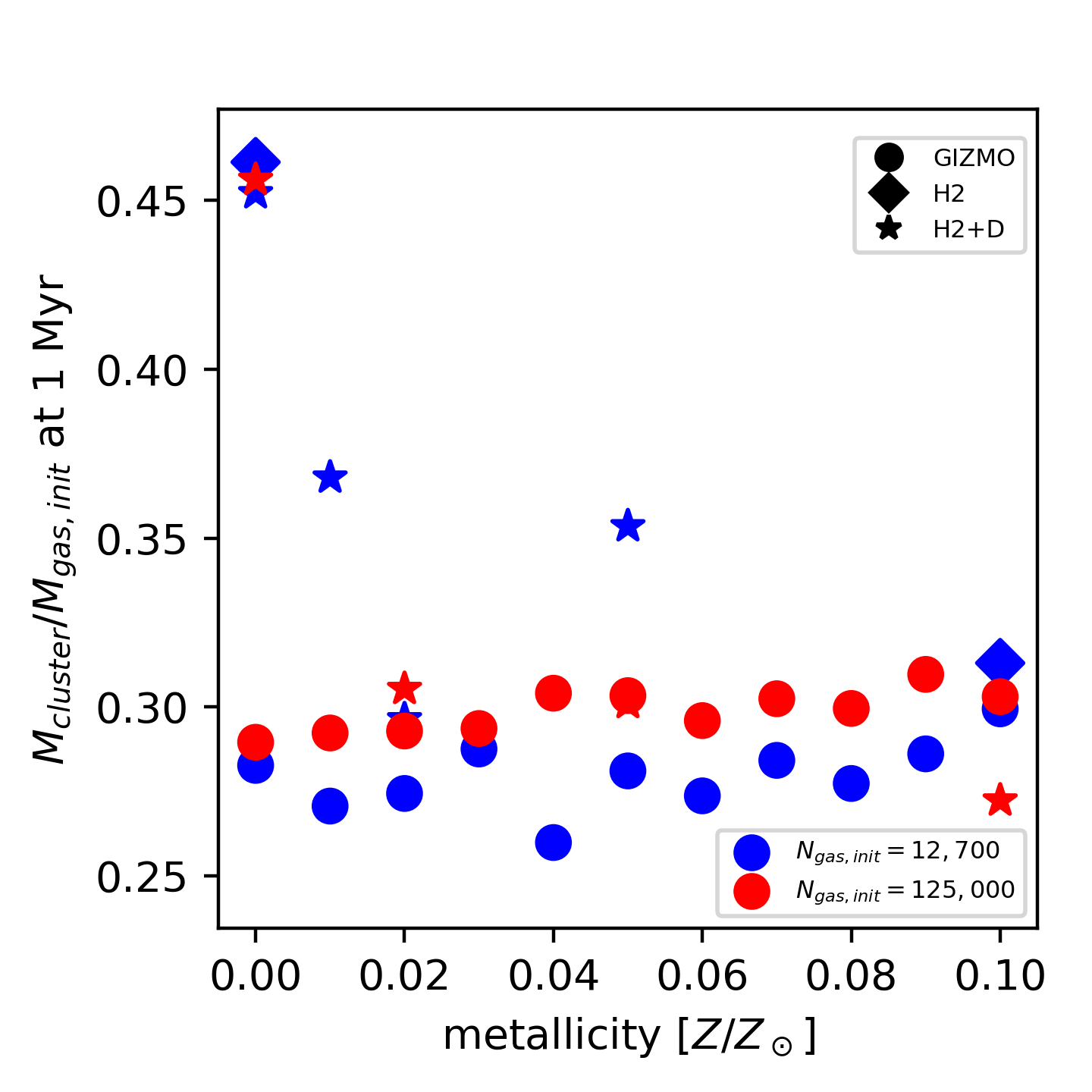}
    \caption{At metallicities of $2\%$ solar and above, metallicity has little effect on fragmentation. Incorporating molecular cooling has a little effect at these metallicities, save for a single low resolution outlier. For models incorporating molecular cooling, for metallicities above $4\%$ solar, increasing resolution results in an increase of fragmentation.}
    \label{fig:vary_metals}
\end{subfigure}

\begin{subfigure}[b]{0.4\textwidth}
	\includegraphics[width=\textwidth]{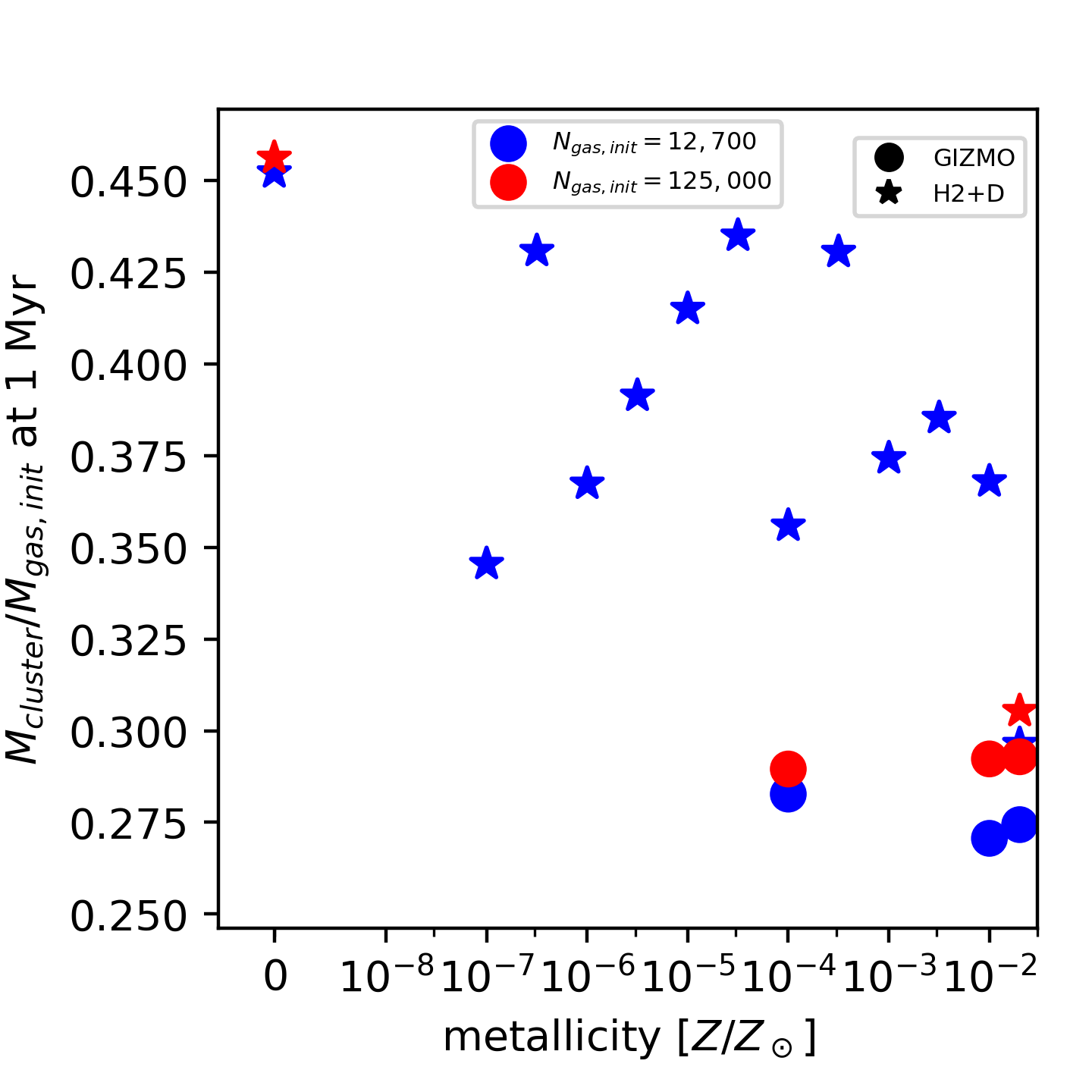}
    \caption{At metallicities of  $2\%$ solar and below and with H2+D cooling, fragmentation is decreased as compared to metallicities above. The transition to zero metallicity is relatively smooth. There is no sharp threshold in any metallicity simulated. At zero metallicity, the differences to the baseline simulation can be substantial.}
    \label{fig:vary_metals_small}
\end{subfigure}
\label{fig:metals}
\end{figure}

\textbf{Figure ~\ref{fig:vary_metals}} shows that at metallicities of $2\%$ solar and above, metallicity has little effect on fragmentation, influencing $M_{cluster}$ by less than $\sim 3\%$ compared to baseline regardless of resolution. Incorporating molecular cooling has a less than $\sim 1\%$ effect on $M_{cluster}$ at these metallicities, save for a single low resolution outlier. The effect of varying metallicity and cooling physics is largely independent of resolution. For GIZMO cooling, increasing resolution reduces fragmentation  independent of metallicity. For models incorporating molecular cooling, increasing resolution results in an increase of fragmentation for metallicities above $4\%$ solar. 

    We perform an additional suite of ultra-low metallicity simulations to investigate the impact of metallicity in our models.  \textbf{Figure ~\ref{fig:vary_metals_small}} shows that for simulations without molecular cooling, metallicity has no effect on fragmentation. Thus, in this section we focus on models which incorporate molecular cooling. For these models, we see that the transition to zero metallicity is relatively smooth. There is no sharp threshold in metallicity at $10^{-3.5} Z/Z_\odot$ which produces a sharp drop in fragmentation as was posited by \cite{bromm_fragmentation_2001}, or, indeed at any other metallicity simulated in the range  $10^{-7}-10^{-2} Z/Z_\odot$. This appears consistent with previous more idealized collapsing-core studies by \cite{jappsen_star_2009} and \cite{safranek-shrader_fragmentation_2010}.
    
At zero metallicity, the differences to the baseline simulation can be substantial. We find differences in the absolute value (percentage wise) of up to $\sim 21-26\%$, but the sign of the percentage different is not consistent with resolution. The choice of a 9 species or a 12 species molecular cooling network has little effect, the difference being less than $2\%$ even for the most extreme case of zero metallicity. For metallicities $1\%$ solar and below, the effect ranges from a drop in fragmentation from the baseline from $\sim 12-25\%$ with the largest drop seen at zero metallicity. 

\begin{figure}
\centering
\begin{subfigure}[b]{0.4\textwidth}
	\includegraphics[width=\columnwidth]{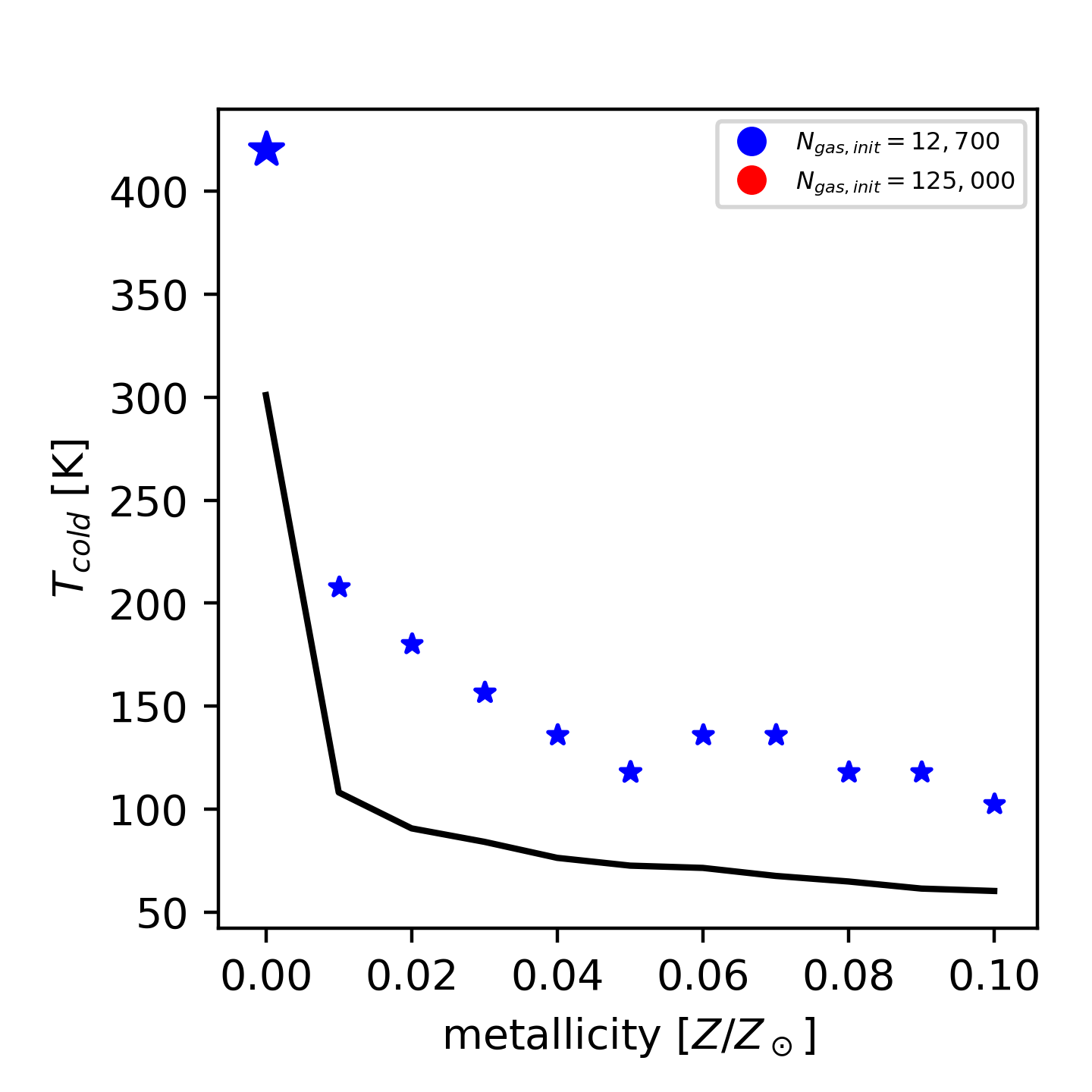}
    \label{fig:temphist_normalmetals}

\end{subfigure}

\begin{subfigure}[b]{0.4\textwidth}
	\includegraphics[width=\columnwidth]{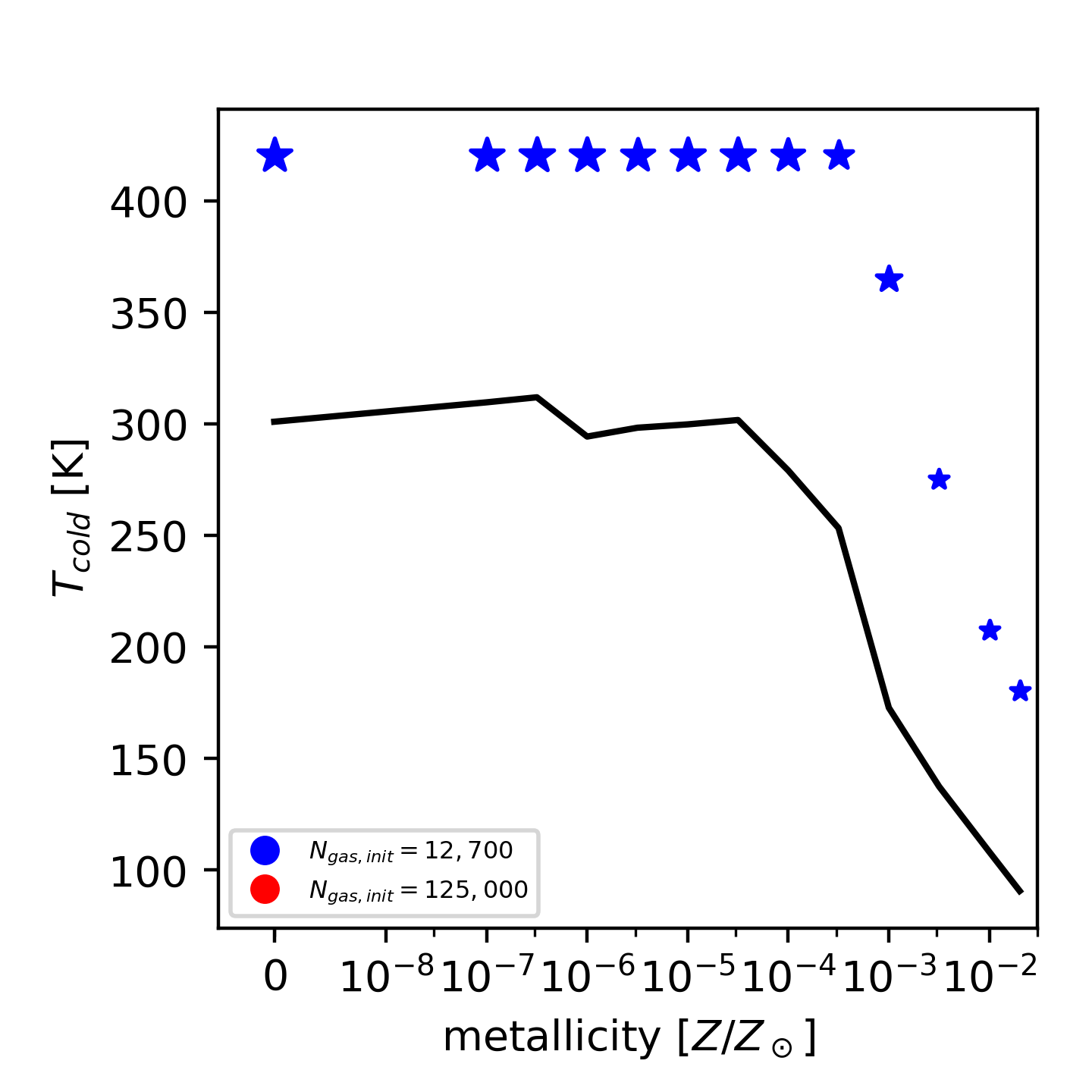}
    \label{fig:temphist_lowmetals}
\end{subfigure}
\caption{The temperature of the cold phase of the gas at at the peak of star formation activity, $T_{cold}$ is plotted for simulations with Grackle H2+D cooling and varied metallicity. The size of the marker is a function of the number of particles at at $T_{cold}$ divided by the maximum $T_{cold}$ seen in any simulation. The black line indicates the temperature where 5\% of gas is below that temperature. In simulations incorporating Grackle cooling, as we go to lower metallicity, $T_{cold}$ scales from $\sim 100K$ to $\sim 450K$, for a turbulent Mach number that goes from $\approx 9$ to $\approx 4$. Thus as metallicity decreases we approach monolithic collapse as compared to runaway fragmentation. Native GIZMO cooling appears to cool more efficiently at all metallicities meaning that this transition is not seen. Grackle results level off below a certain metallicity because the pure molecular cooling begins to dominate over dust cooling.}\label{fig:temphist}
\end{figure}

The average star formation time (median time when the stars formed, from the start of the simulation) was approximately 0.2  Myr. The surface density  of the gas at 0.2 Myr is $\sim 2 \times 10^4 M_\odot/pc^2$; given this high density most of the gas is self-shielding and cold. To determine the effect of metallicity on fragmentation, we examined the distribution of the gas temperature at 0.2 Myr. We computed a histogram with 50 bins logarithmic in temperature, and denoted the average temperature of the bin which contained the majority of the particles as the temperature of the cold phase of the gas $T_{cold}$. We then plotted this temperature as a function of metallicity in Fig.~\ref{fig:temphist}. In this figure, the size of the marker is a function of the number of particles at at $T_{cold}$ divided by the maximum $T_{cold}$ seen in any simulation. 

In simulations incorporating Grackle cooling, as we go to lower metallicity, $T_{cold}$ scales from $\sim 100K$ to $\sim 450K$, for a turbulent Mach number that goes from $\approx 9$ to $\approx 4$. This alters the physical nature of fragmentation.  The transition from a runaway fragmentation cascade to monolithic collapse happens at Mach $\approx 3$ (David Guszejnov, in preparation). Mach 3.3 is also the number associated with the Larson-Penston isothermal collapse solution \citep{larson_numerical_1969,penston_dynamics_1969}. Thus, when $T_{cold}$ approaches 500K, as metallicity decreases, we approach monolithic collapse as compared to runaway fragmentation. Native GIZMO cooling appears to cool more efficiently at all metallicities (likely due to the lookup tables used extrapolating metal-line cooling more efficiently than should be done, as they are calibrated at $\gtrsim 10^{-2} Z_\odot$, with $T_{cold}$ being in the range of $25-44K$, meaning that this transition is not seen with Native GIZMO cooling. 

\begin{figure}
\centering
	\includegraphics[width=0.4\textwidth]{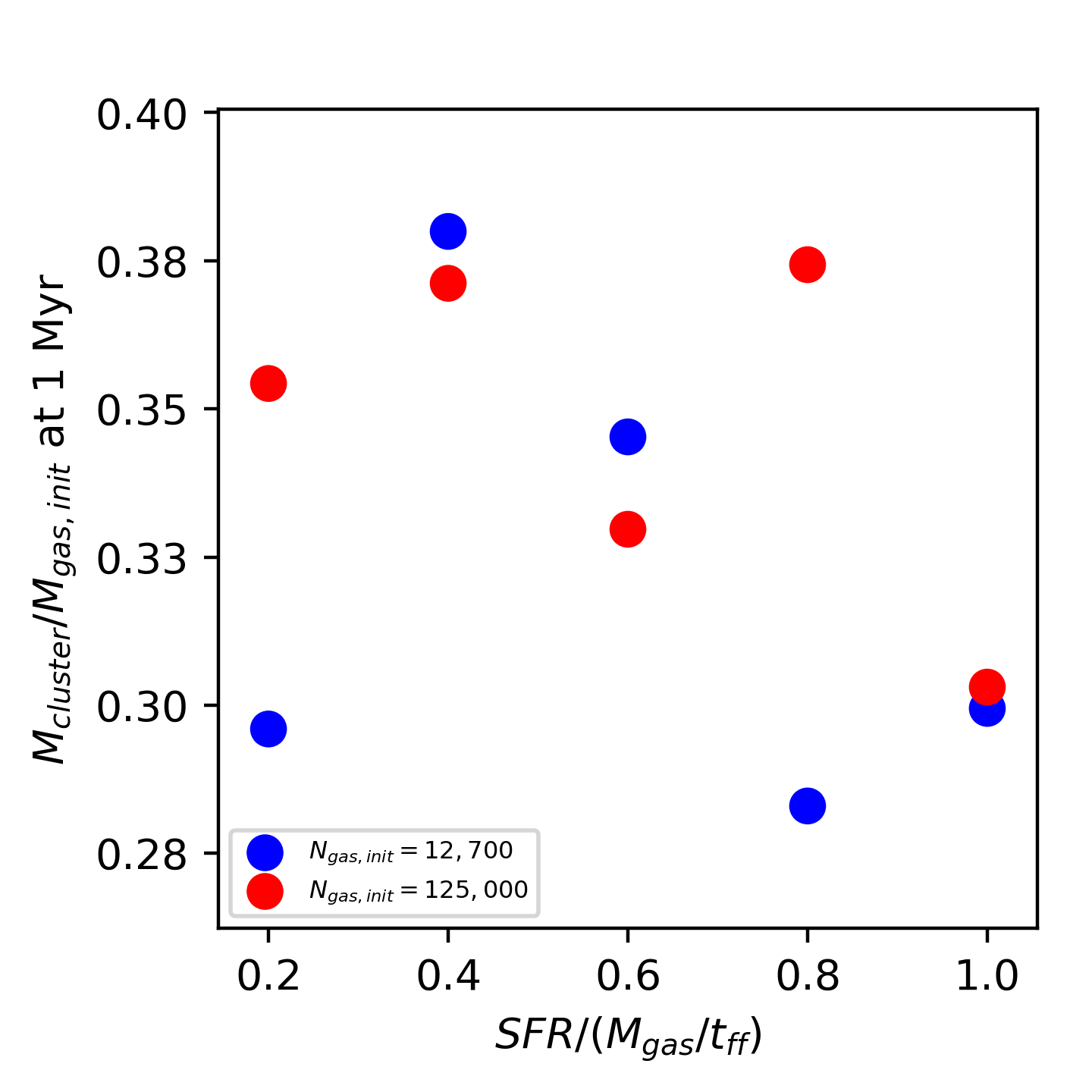}
    \caption{The mass of the most massive cluster of star particles at 1 Myr $M_{cluster}$, is normalized by the initial mass of the gas, $M_{gas,init}$ and plotted as a function of numerical parameter of the local star formation rate per effective free fall time,  $\epsilon_{ff}$.  Simulations with GIZMO cooling, fixed metallicity of $0.1 Z/Z_\odot$ and $\lfloor T \rfloor=10$K.  For fixed metallicity, varying the numerical parameter $\epsilon_{ff}$ has little effect on fragmentation. For those simulations where resolution has any appreciable effect, increasing resolution decreases fragmentation slightly.}
    \label{fig:vary_sfeff}
\end{figure}

The numerical parameter of the local star formation rate per effective free fall time, $\epsilon_{ff}$ is referenced as $\frac{SFR}{M_{gas}/t_{freefall}}$ in this work. Fig.~\ref{fig:vary_sfeff} shows that for fixed metallicity, varying the numerical parameter $\epsilon_{ff}$ has little effect on fragmentation (consistent with scale results \citep{orr_what_2017}). The effect ranges from $\sim -1-15\%$ as compared to an $\epsilon_{ff}=1.0$, but the magnitude of that percentage difference is inconsistent. For those simulations where resolution has any appreciable effect, increasing resolution decreases fragmentation slightly. 

Some numerical studies have argued that in the presence of an intense Lyman-Werner flux inhibits cooling, causing a cloud of T $\sim 8000$ K to collapse with little fragmentation \citep{bromm_formation_2003,wise_resolving_2008, regan_pathways_2009,latif_black_2013,prieto_gas_2013,regan_numerical_2014,latif_magnetic_2014,choi_supermassive_2015,becerra_formation_2015}. In this paper, we use the numerical gas temperature floor $\lfloor T \rfloor$ as a proxy for a Lyman-Werner background of varying intensity (or any other physics that suppresses cooling). The higher $\lfloor T \rfloor$, the more intense the background.  

\begin{figure}
\centering
	\includegraphics[width=0.4\textwidth]{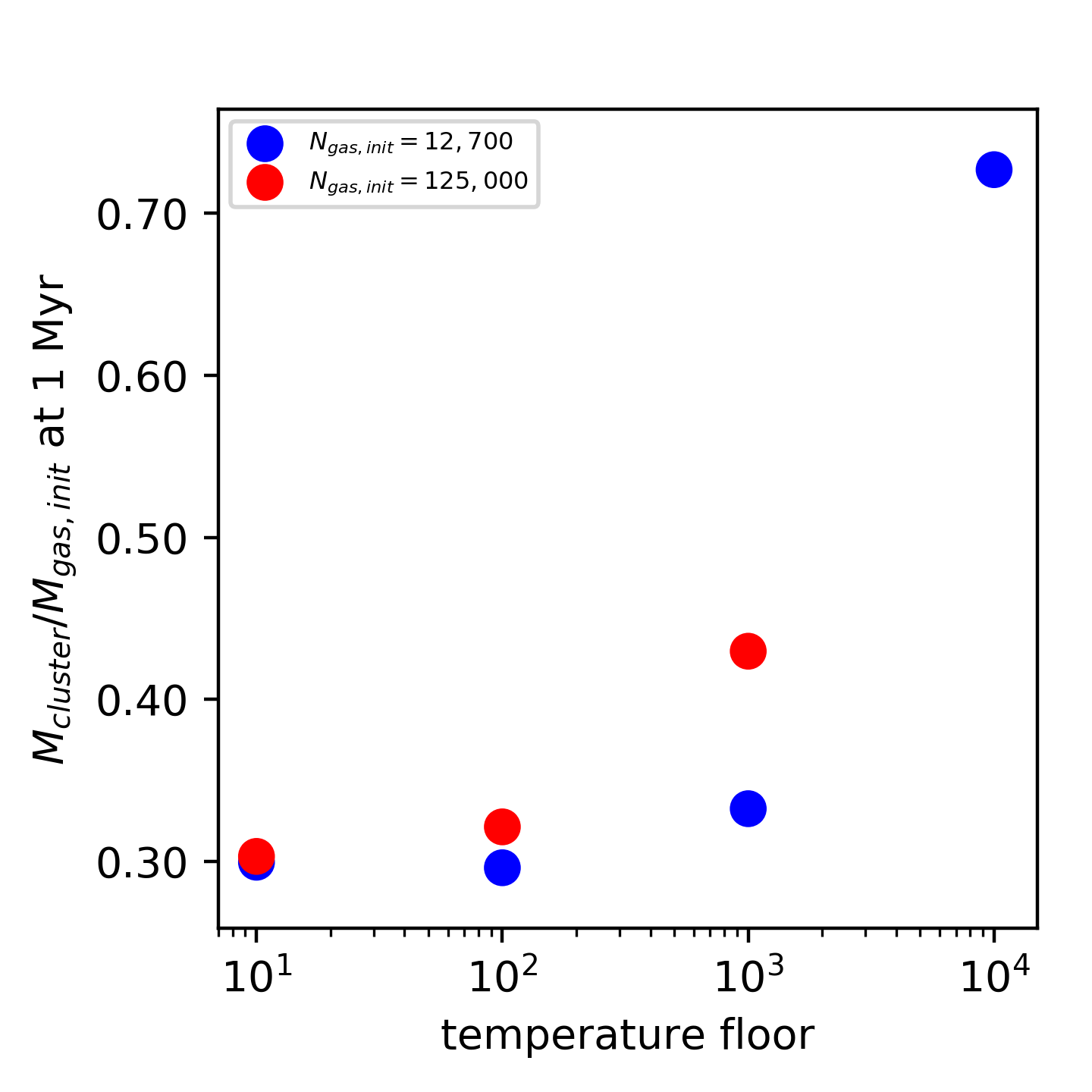}
    \caption{ 
    The mass of the most massive cluster of star particles at 1 Myr $M_{cluster}$ is normalized by the initial mass of the gas, $M_{gas,init}$ and plotted as a function of gas temperature floor  $\lfloor T \rfloor$. Simulations use GIZMO cooling and fixed metallicity of $0.1 Z/Z_\odot$ and $\epsilon_{ff}=1.0$. The mass of the most massive cluster increases as the gas temperature floor  $\lfloor T \rfloor$  is increased, as is expected theoretically.  There is only a slight reduction in fragmentation between a $\lfloor T \rfloor$ of $10^1$K and one of $10^{2}$ K. For a floor of $10^3$K and above, the reduction becomes appreciable. We note that star formation histories of these simulations show that star formation has plateaued at 1 Myr.}
    \label{fig:vary_mingas}
\end{figure}

Fig.~\ref{fig:vary_mingas} shows that that the mass of the most massive cluster increases as the gas temperature floor  $\lfloor T \rfloor$  is increased, as is expected theoretically.  There is only a slight reduction in fragmentation between a $\lfloor T \rfloor$ of $10^1$K and one of $10^{2}$ K. For a floor of $10^3$K and above, the reduction becomes appreciable, $\sim 20\%$ at $10^3$ K and $\sim 60\%$ at $10^4$ K.  In the $10^4$ K run, the hotter gas suppresses formation, encouraging the formation of a single massive cluster. Then, feedback from this single cluster disrupts the cloud more efficiently than it would with many smaller clusters, as there is no cancellation of gas momentum from multiple sources. 

We also investigated an alternative proxy of fragmentation, the fraction of gas which turns into stars by 1 Myr i.e. the mass in stars at 1 Myr compared to the initial gas mass $M_{stars}/M_{gas,init}$. In this proxy, the higher the fraction, the greater the fragmentation. Our results using this definition were qualitatively similar and presented no additional conclusions.

\section{CONCLUSIONS}
\label{sec:conclusions}
We performed a detailed suite of 3D hydrodynamics simulations using the GIZMO code on turbulent primordial gas clouds. We explore the effect of assumptions as to metallicity, cooling physics, numerical star formation parameters, and gas temperature floor on fragmentation.

At fixed local star formation rate per free fall time and minimum gas temperature, the detailed cooling physics, metallicity choice, and resolution probed do not have a strong effect on fragmentation for metallicities above $2\%$ solar. The same is true of metallicities below $2\%$ solar in scenarios without molecular cooling. If molecular cooling is incorporated, metallicities below $2\%$ solar show a decrease in fragmentation. 

At fixed metallicity, gas temperature floor and cooling physics, the numerical parameter of local star formation rate per effective free fall time has a small and inconsistent effect. At fixed local star formation rate per effective free fall time, metallicity and cooling physics, the gas temperature floor has no effect below $10^3$ K. At a value of $10^3$ K the effect is on the order of choosing a metallicity below $1\%$ solar, At $10^4$ K the effect can be up to a $\sim 60\%$ reduction in fragmentation. 

There is a large of work at metallicities $\gtrsim 10^{-4}-10^{-2} Z_\odot$ (see \cite{offner_origin_2014} for a review) showing that the IMF is nearly metallicity-independent (i.e. there is no excess of higher-mass stars at lower metallicity, within this range). Our work is consistent with this picture, extending it to still lower metallicity. As we have a fixed fixed star particle mass, we do not simulate accretion onto a protostar, nor are our simulations suitable for an IMF study. We assume main sequence stars, where as over time these scales the stars may be protostars so we would need a protostar evolution code complete with accretion to self-consistently model the IMF. Some work shows this accretion could provide feedback via radiative heating of dust halting further fragmentation below $\lesssim 0.1 M_\odot$, however this is not expected to be relevant at the lower metallicities simulated (nor at the mass scales of interest for direct-collapse black holes). Our intention instead here is simply to gain some intuition for how fragmentation and star formation proceeds with more detailed metallicity-dependent cooling physics at much larger mass scales relevant for direct-collapse black holes. Future work will include higher resolution simulations of a non-fixed star particle mass incorporating a self-consistent model of the Lyman-Werner UV background.

We conclude: 
\begin{enumerate}
\item There is no cutoff metallicity after which there is a sharp decrease in fragmentation under the conditions that we have simulated.
\item  The numerical gas temperature floor parameter has the largest potential impact on fragmentation 
\item Fragmentation decreases with decreasing metallicity as $T_{cold}$ increases, eventually crossing the transition Mach number of $\approx 3$ changing from runaway fragmentation cascade to monolithic collapse. Metallicities below $0.05 Z/Z_\odot$ showing this effect.
\item Clouds below $2\%$ solar metallicity in simulations incorporating molecular cooling show a decrease in fragmentation
\end{enumerate}

\section*{Acknowledgements}
CCM was supported by the NSF Astronomy and Astrophysics Postdoctoral Fellowship under award AST-1501208. Support for PFH was provided by an Alfred P. Sloan Research Fellowship, NSF Collaborative Research Grant \#1715847 and CAREER grant \#1455342. Numerical calculations were run on the Caltech compute cluster ``Wheeler,'' allocations from XSEDE TG-AST130039 and PRAC NSF.1713353 supported by the NSF, and NASA HEC SMD-16-7592.



\bibliographystyle{mnras}
\bibliography{Zotero} 



\bsp	
\label{lastpage}
\end{document}